\documentclass{aa}
\usepackage{amssymb,amsmath}
\usepackage{graphicx}
\usepackage{xcolor,natbib}
\usepackage{multirow}
\usepackage[T1]{fontenc}
\usepackage{ae,aecompl}
\usepackage{newtxtext, newtxmath}
\usepackage{float}
\usepackage{subfigure}
\usepackage{longtable}
\usepackage{enumitem}
\usepackage{longtable,listings}
\usepackage[flushleft]{threeparttable}
\usepackage{parcolumns}
\usepackage{hyperref}

\def\n2{[N~{\sc ii}]$\lambda6583$\AA}

\def\o3{[O~{\sc iii}]$\lambda4364$\AA}
\def\obj{SDSS J1251+0613}

\begin{document}

\title{No apparent forbidden/permitted narrow emission lines in the broad line quasar SDSS J1251+0613}

\titlerunning{Evolving AGN NLRs}


\author{XueGuang Zhang}

\institute{Guangxi Key Laboratory for Relativistic Astrophysics, School of Physical Science and Technology, GuangXi University, 
No. 100, Daxue Road, Nanning, 530004, P. R. China 
\ \ \ \email{xgzhang@gxu.edu.cn}}

\abstract
{Strong both broad and narrow emission lines from central broad emission line regions (BLRs) and narrow emission line regions 
(NLRs) are fundamental spectroscopic characteristics of broad line Active Galactic Nuclei (BLAGNs). Lack of central BLRs leads 
to the identified true Type-2 AGNs without central hidden BLRs, providing clues on formation/suppression of AGN BLRs. Whether 
were there BLAGNs with lack of central NLRs is still an open question. Here, in the blue quasar SDSS J1251+0613, both blue 
continuum emissions and broad emission lines can be clearly detected in its SDSS spectrum, however there are no apparently 
detected narrow emission lines in the optical/NUV bands, leading to no central normal NLRs in the blue quasar SDSS J1251+0613. 
In order to explain the lack of NLRs, evolving NLRs is proposed that the radial outflows are carrying materials from BLRs to 
NLRs and the current narrow line emission materials are lying closer to the outer side of central BLRs in SDSS J1521+0613. The 
results in this manuscript can indicate a new unique subclass of BLAGNs, the BLAGNs without central normal NLRs, which will 
provide potential clues on physical origin/evolution of AGN NLRs. 
}


\maketitle

\section{Introduction}

	Broad and narrow emission lines can be clearly detected in the optical band and NUV (near ultraviolet) band of broad line 
Active Galactic Nuclei (BLAGNs), related to central two fundamental structures of broad emission line regions (BLRs) and narrow 
emission line regions (NLRs), as the apparent spectroscopic features in the composite spectrum of BLAGNs reveal 
\citep{zk97, bt01, vr01, ss12, sf16, pw17, ks21, tl24}. Meanwhile, when measuring emission lines in BLAGNs, the 
emission lines are also considering two source contributions, one source contribution for components from BLRs and the other one 
for components from NLRs, such as the discussions in \citet{sm00, pf04, gh05, bw09, sr11, pw12, oy15, wy15, ll19, zh21, zh22, 
zh23, wg24, zh24, zm25}. Therefore, both broad emission lines from BLRs and narrow emission lines from NLRs are fundamental 
optical/NUV spectroscopic characteristics of BLAGNs.

	Until now, there is basic structure information of central BLRs and NLRs, such as their distances to central BHs as 
reported in \citet{ks00, ben13, liu13, dz18, zh24b}. Unfortunately, there are no clear reports on physical materials origin in 
BLRs or in NLRs, besides some potential connections between BLRs and NLRs through radial outflows, such as the discussions in 
\citet{fc13, dw14, fk18, tk21, zh21, mh22, cr24, sr24, ht25}. Furthermore, among studies of AGNs, there is one special subclass 
of AGNs, the true Type-2 AGNs without central hidden BLRs, such as the cases in \citet{tr01, tr03, hm04, hk12, zh14, pw16, zh22b}. 
The study of the properties of true Type-2 AGNs can provide clues on formation/suppression of central BLRs probably 
depending on central AGN activities, as discussed in \citet{eh09, ca10, ip15, en16}. Similarly, to detect and study BLAGNs without 
NLRs could probably provide some clues on formation/suppression of AGN NLRs. Unfortunately, there is no one report on BLAGNs 
without NLRs. Therefore, to detect and report such a BLAGN without NLRs is the main objective of the manuscript.

	The manuscript is organized as follows. Section 2 presents the spectroscopic results and main discussions of the BLAGN 
SDSS J125157.91+061341.63 (=SDSS J1251+0613) at redshift around 0.375, to confirm normal apparent broad emission lines but no 
apparent reliable narrow emission lines. Section 3 shows the necessary discussions. The main summary and conclusions are given 
in Section 4. Throughout the manuscript, we have adopted the cosmological parameters of $H_{0}$=70 km s$^{-1}$ Mpc$^{-1}$, 
$\Omega_{m}$=0.3, and $\Omega_{\Lambda}$=0.7.

\begin{figure*}
\centering\includegraphics[width = 18cm,height=8cm]{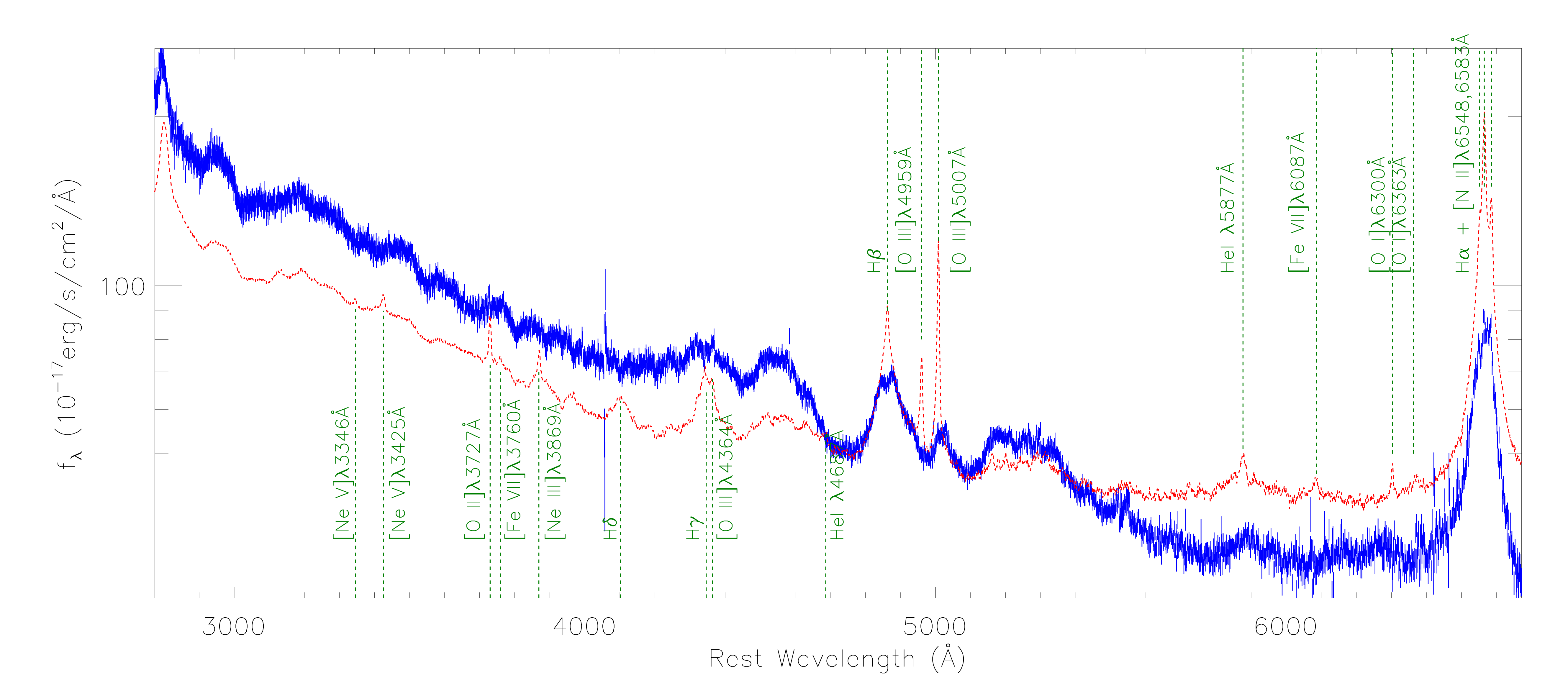}
\caption{The SDSS spectrum of SDSS J1251+0613 (blue dots plus error bars) and the composite spectrum of SDSS quasars (red dashed 
line) from \citet{vr01}. Here, the composite spectrum is scaled with continuum intensity about 
$52\times{10}^{-17}{\rm erg/s/cm^2/\textsc{\AA}}$ at 5000\AA. The vertical dashed lines in dark green mark positions of the 
commonly apparent and strong both forbidden and permitted narrow emission lines in blue quasars. 
}
\label{spec}
\end{figure*}

\begin{figure*}
\centering\includegraphics[width = 18cm,height=10cm]{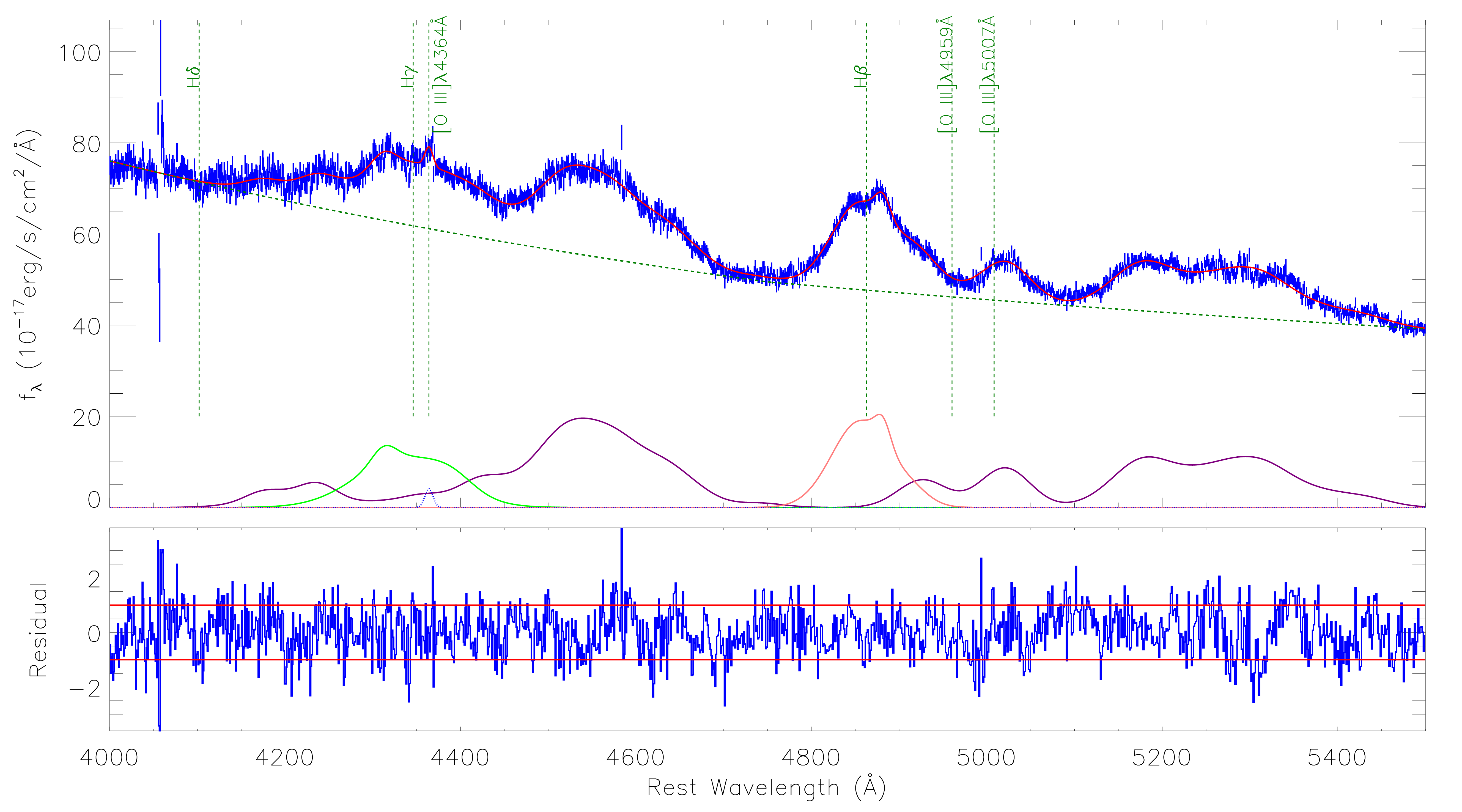}
\caption{The best fitting results (solid red line in top panel) and the corresponding residuals (bottom panel) to the emission 
lines (blue dots plus error bars) of SDSS J1251+0613 with rest wavelength from 4000\AA~ to 5500\AA. In the top panel, dashed line 
in dark green shows the determined power law AGN continuum emissions, solid lines in green and in pink show the determined broad 
H$\gamma$ and broad H$\beta$, solid purple line shows the determined optical Fe~{\sc ii} emissions, dotted blue line shows the 
determined probable [O~{\sc iii}]$\lambda$4364\AA~ emission component. And the vertical dashed lines in dark green mark the 
positions of the commonly apparent and strong narrow emission lines in blue quasars. In the bottom panel, horizontal solid red 
lines show residuals=$\pm1$.}
\label{hb}
\end{figure*}

\begin{figure}
\centering\includegraphics[width = 8cm,height=5cm]{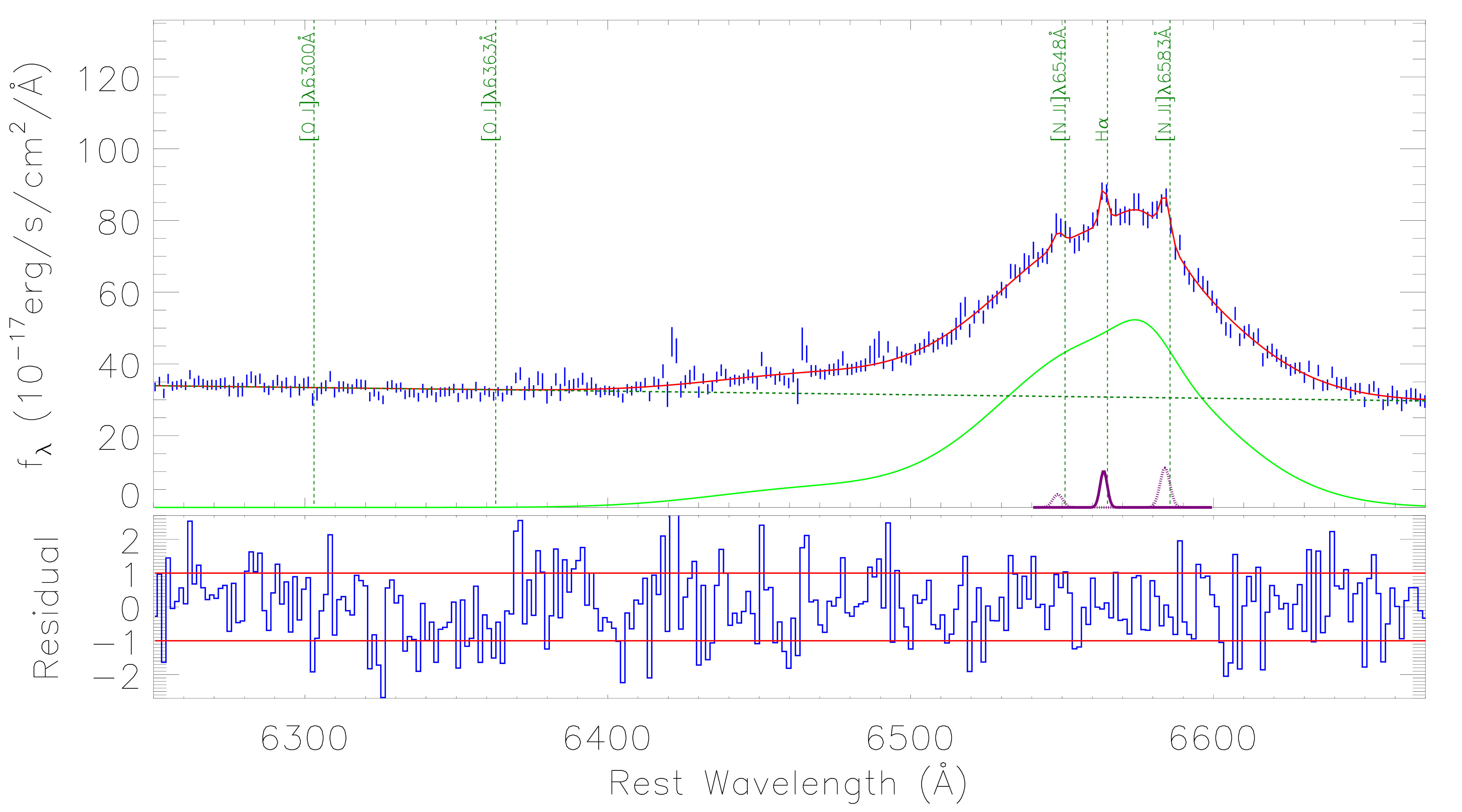}
\caption{The best fitting results (solid red line in top panel) and the corresponding residuals (bottom panel) to the emission 
lines (blue dots plus error bars) of SDSS J1251+0613 with rest wavelength from 6250\AA~ to 6670\AA. In the top panel, dashed 
line in dark green shows the determined power law AGN continuum emissions, solid green line and solid purple line shows the 
determined broad and probable narrow components in H$\alpha$, dotted purple lines show the determined components in [N~{\sc ii}] 
doublet. And the vertical dashed lines in dark green mark the positions of the commonly apparent and strong narrow emission 
lines in blue quasars. In the bottom panel, horizontal solid red lines show residuals=$\pm1$.}
\label{ha}
\end{figure}

\begin{figure}
\centering\includegraphics[width = 9cm,height=5cm]{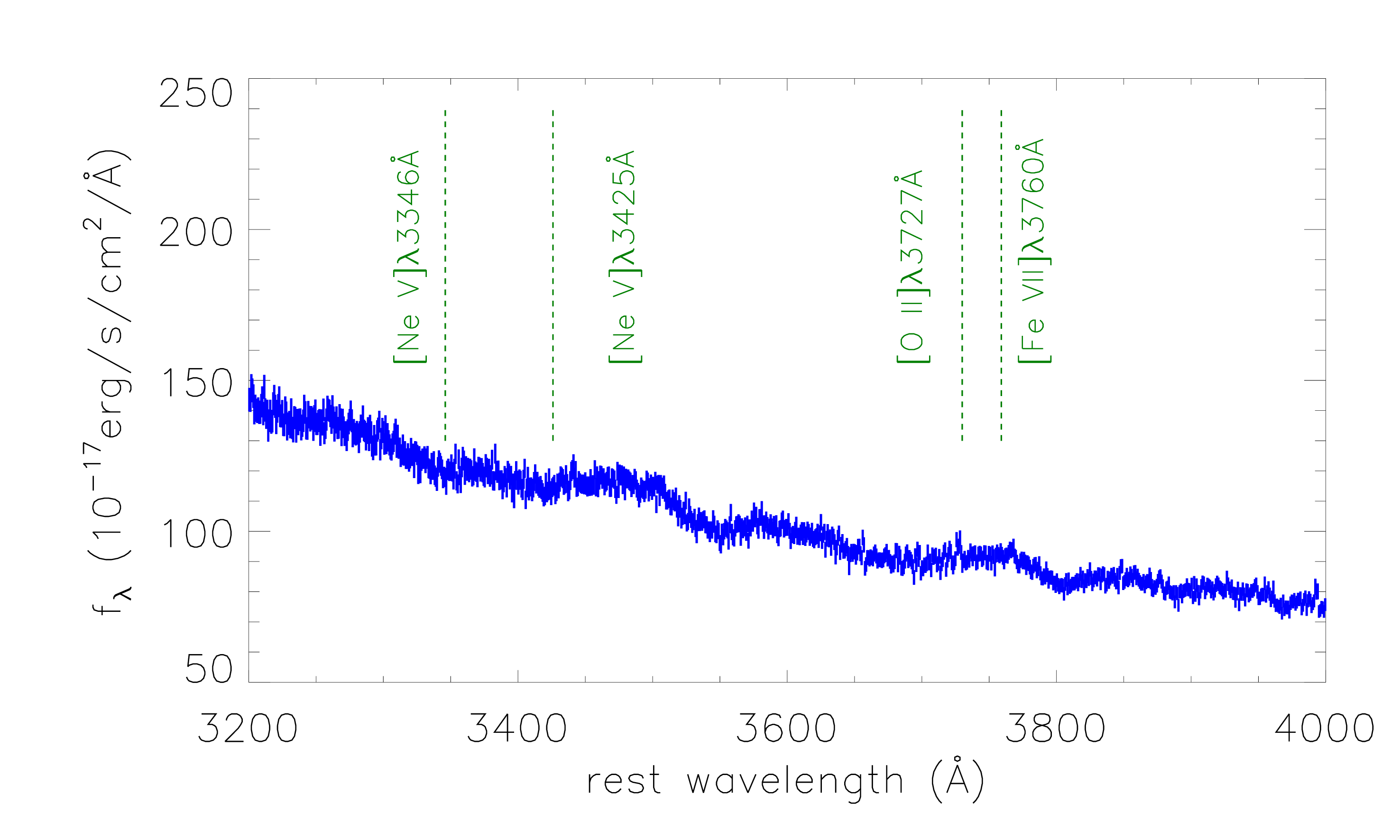}
\caption{Spectroscopic properties with rest wavelength from 3200\AA~ to 4000\AA. The vertical dashed lines in dark green mark 
the positions of the commonly apparent and strong narrow emission lines in blue quasars.}
\label{mgii}
\end{figure}

\section{Spectroscopic results of SDSS J1251+0613}

\begin{table}
\caption{Parameters of each emission component}
\begin{tabular}{llll}
\hline\hline
	line & $\lambda_0$ & $\sigma$ & flux \\
\hline\hline
	H$\beta_1$ & 4855.9$\pm$3.1 & 2024$\pm$117  & 1560$\pm$136 \\
	H$\beta_2$ & 4881.5$\pm$1.4 & 543$\pm$111  & 114$\pm$45 \\
	H$\beta_3$ & 4914.6$\pm$10.2 & 1278$\pm$457   & 132$\pm$103 \\
	H$\gamma_1$ & 4312.8$\pm$2.8 & 954$\pm$267 & 114$\pm$61   \\
	H$\gamma_2$ & 4334.8$\pm$8.6 & 3649$\pm$297 & 1453$\pm$262   \\
	H$\gamma_3$ & 4391.5$\pm$11.6 & 1665$\pm$1057 & 150$\pm$212   \\
	\o3 & 4364.7$\pm$0.9  & 302$\pm$82  &  46$\pm$16 \\
	Fe~{\sc ii} &  134$\pm$4* & 1710$\pm$40  &  7262$\pm$540  \\
	H$\alpha_1$ &  6563.8$\pm$0.9 & 1568$\pm$411 & 3961$\pm$92 \\
	H$\alpha_2$ &  6468.4$\pm$6.7 & 1540$\pm$224 & 437$\pm$79 \\
	H$\alpha_3$ &  6577.6$\pm$1.6 & 398$\pm$77 & 195$\pm$52 \\
	H$\alpha_4$ &  6563.7$\pm$0.3 & 55$\pm$18 & 30$\pm$10 \\
	\n2 & 6584.0$\pm$0.4 & 73$\pm$18 & 45$\pm$13 \\
\hline\hline
	\multicolumn{4}{c}{Disk parameters for the broad H$\beta$} \\
	\multicolumn{4}{c}{$r_0=96\pm27$,~$r_1=2142\pm630$,~$i=61\pm17\degr$}\\
	\multicolumn{4}{c}{$q=2.02\pm0.06$,~$e=0.87\pm0.04$}\\
	\multicolumn{4}{c}{$\sigma_L=500\pm66{\rm km/s}$,~$\phi_0=-3.5\pm1.2\degr$} \\
\hline\hline
\end{tabular}\\
Notice: The first column shows the determined emission component. H$\beta_1$, H$\beta_2$, H$\beta_3$ and H$\gamma_1$, H$\gamma_2$, 
H$\gamma_3$ are the determined three Gaussian components in H$\beta$ and in H$\gamma$, and H$\alpha_1$, H$\alpha_2$, H$\alpha_3$, 
H$\alpha_4$ are the determined four Gaussian components in H$\alpha$ (three broad components plus one probable narrow component). 
The second column, the third column and the fourth column show the determined central wavelength in units of \AA, the second moment 
in units of km/s, and the emission flux in units of $10^{-17}{\rm erg/s/cm^2}$. \\
For the optical Fe~{\sc ii} emissions, the shifted velocity and the broadened velocity in units of km/s are listed in the second 
column and the third column, and the total flux is listed in the fourth column. \\
The second part of the table lists the model parameters for the elliptical accretion disk model applied to describe the asymmetric 
broad H$\beta$ as shown in Fig.~\ref{dbp}.
\end{table}

	SDSS J1251+0613 with plate-mjd-fiberid=1792-54270-0260 is a SDSS pipeline classified blue quasar at redshift around 0.375 
in SDSS DR16 (Sloan Digital Sky Survey, Data Release 16) \citep{ss13, as15}. The high-quality SDSS spectrum (median signal to noise 
ratio about 37) with exposure time about 3000 seconds is shown in Fig.~\ref{spec} with apparent blue quasar-like continuum emissions 
and apparent broad emission lines. Meanwhile, comparisons are also shown in Fig.~\ref{spec} between the spectrum of SDSS J1251+0613 
and the composite spectrum of SDSS quasars in \citet{vr01}. It is clear that SDSS J1251+0613 has the normal spectral energy 
distributions (continuum emissions) similar as the other blue quasars in SDSS. However, from the comparisons, there are no expected 
strong forbidden narrow emission lines of [O~{\sc iii}], [O~{\sc ii}], [O~{\sc i}], [Fe~{\sc vii}], [Ne~{\sc v}], [Ne~{\sc iii}], 
[N~{\sc ii}] or expected strong permitted narrow Balmer lines and permitted narrow He~{\sc i} lines which are common in quasars but 
cannot be found in the spectrum of SDSS J1251+0613. In order to get clearer information of probable narrow emission lines in SDSS 
J1251+0613, the following emission line fitting procedures are applied.

	For the emission lines with the rest wavelength from 4000\AA~ to 5500\AA, the following model functions have been 
employed. Four Gaussian functions are applied to describe each Balmer emission line (H$\beta$, H$\gamma$). Two Gaussian 
functions are applied to describe the core and the extended components in each line in [O~{\sc iii}]$\lambda4959, 5007$\AA~ doublet and 
[O~{\sc iii}]$\lambda4364$\AA. Two Gaussian functions are applied to describe probable He~{\sc ii}$\lambda4687$\AA~line. The shifted, 
strengthened and broadened optical Fe~{\sc ii} templates in \citet{bg92, kp10} are accepted to describe the optical Fe~{\sc ii} 
emission features. A Power law function is applied to describe the continuum emissions underneath the emission lines. 
When the model functions above are applied, there are no restrictions on the model parameters, except the emission intensity of each 
Gaussian component not smaller than zero and that the [O~{\sc iii}] doublet have the same line widths, the same redshift and the 
fixed flux ratio to be 3. Then, through the Levenberg-Marquardt least-squares minimization technique (the known MPFIT 
package) \citep{mc09}, the best descriptions and the corresponding residuals can be determined to the emission lines, as shown in 
Fig.~\ref{hb} with the corresponding $\chi^2/dof=1037.1/1343\sim0.77$ ($dof$ as the degree of freedom). Residuals are calculated by 
the SDSS spectrum minus the best fitting results and then divided by the uncertainties of the SDSS spectrum. The determined model 
parameters of the emission components are listed in Table~1. Here, three additional points should be noted. First, Gaussian functions 
have been applied to describe probable [O~{\sc iii}]$\lambda4959, 5007$\AA~ doublet and He~{\sc ii}, however, there are no reliable 
determined model parameters, due to the determined model parameters smaller than their uncertainties (emission intensity near to zero). 
Therefore, there are no parameters listed in Table~1 for the [O~{\sc iii}]$\lambda4959, 5007$\AA~ doublet and He~{\sc ii}. Second, 
four Gaussian functions are applied to describe H$\beta$ (H$\gamma$), however, only three Gaussian components have their determined 
model parameters larger than their corresponding uncertainties, therefore, there are only three components with their model parameters 
listed in Table~1 for H$\beta$ and H$\gamma$. Third, the flux ratio about 1.05 of H$\beta$ to H$\gamma$ is very different from the 
determined value 3.3 from the composite spectrum of SDSS quasars in \citet{vr01}, probably due to the determined 
H$\gamma$ including contributions of broad Fe~{\sc ii} component around 4351\AA~\citep{sp03}. Due to no effects of determined 
H$\gamma$ on the measurements of narrow emission lines, there are no further discussions on broad emission components around H$\gamma$.

	Moreover, similar fitting procedure has been applied to describe the emission lines around H$\alpha$ with the rest wavelength 
from 6250\AA~ to 6670\AA~(the maximum wavelength covered by the SDSS in \obj). Four Gaussian functions are applied to describe 
H$\alpha$. Two Gaussian functions are applied to describe [N~{\sc ii}]$\lambda6548,6583$\AA~ doublet, and also applied to 
describe the [O~{\sc i}]$\lambda6300,6363$\AA~ doublet. Power law function is applied to describe the continuum emissions underneath 
the emission lines. When the model functions above are applied, there are no restrictions on the model parameters, except the emission 
intensity of each Gaussian component not smaller than zero and except that the [N~{\sc ii}] doublet have the same line 
widths, the same redshift and the fixed flux ratio to be 3. Then, through the Levenberg-Marquardt least-squares minimization 
technique, the best descriptions and the corresponding residuals can be determined to the emission lines, as shown in Fig.~\ref{ha} 
with the corresponding $\chi^2/dof=294.7/267\sim1.10$. The determined model parameters of the emission components are listed in 
Table~1. And, Gaussian functions have been applied to describe probable [O~{\sc i}]$\lambda6300, 6363$\AA~ doublet, however, there 
are no reliable determined model parameters, due to the determined emission intensity near to zero. Therefore, there are no parameters 
listed in Table~1 for the [O~{\sc i}]$\lambda6300, 6363$\AA~ doublet.

	Furthermore, spectroscopic features around [O~{\sc ii}]$\lambda3727$\AA~ with rest wavelength from 3200\AA~ to 4000\AA~ 
are also shown in Fig.~\ref{mgii}. Apparently, there are no narrow emission lines. Certainly, Gaussian functions have been tried 
to be applied to describe the probable narrow emission lines, however, no reliable narrow Gaussian components can be found. 
Therefore, compared with the composite spectrum of SDSS quasars, there are no detected narrow emission lines of 
[Ne~{\sc v}]$\lambda3346$\AA, [Ne~{\sc v}]$\lambda3425$\AA, [O~{\sc ii}]$\lambda3727$\AA, [Fe~{\sc vii}]$\lambda3760$\AA, and 
[Ne~{\sc iii}]$\lambda3869$\AA~ in \obj.

	Therefore, based on the careful analysis of spectroscopic emission lines, at least three points can be found in \obj. First, 
combined with the quasar-like blue continuum emissions, there are apparent broad emission lines, especially broad Balmer and 
Mg~{\sc ii} emission lines. Second, considering measured line parameters at least five times larger than their uncertainties, there 
are no apparent narrow emission lines in optical/NUV band. Third, if accepted measured line parameters at least two times larger 
than their uncertainties, weak [O~{\sc iii}]$\lambda4364$\AA, [N~{\sc ii}]$\lambda6548,6583$\AA, and narrow H$\alpha$ (the listed 
H$\alpha_4$ in Table~1) could be possibly detected. The results above can indicate that SDSS J1251+0613 is so-far the first unique 
blue quasar without clearly detected forbidden/permitted narrow emission lines in optical/NUV band. Based on the Unified Model 
\citep{ar93, nh15} for different kinds of AGNs, it can be well expected for non-detected broad emission lines in observed spectra 
of Type-2 AGNs, due to serious obscuration of central dust torus on central BLRs. However, there is no way to expect non-detected 
narrow emission lines in BLAGNs.

	However, very higher electron densities than the corresponding critical densities can lead to very weak forbidden narrow 
emission lines, due to efficient suppressions by the collisional de-excitations. Therefore, combining with higher electron densities 
in NLRs, a more direct and natural explanation to the non-detected forbidden/permitted narrow emission lines could be on kinematic 
properties of the NLRs in SDSS J1251+0613. In other words, we accept the assumption that the expected NLRs were very nearer to the 
central BLRs in SDSS J1251+0613. The assumption can lead to non-detected permitted narrow emission lines, because the location near 
to the BLRs leading to broad rather than narrow permitted emission lines including most part of the emissions. And the assumption 
can also lead to non-detected forbidden narrow emission lines, because of the higher electron densities in the clouds near to the 
BLRs in SDSS J1251+0613.

\begin{figure*}
\centering\includegraphics[width = 18cm,height=10cm]{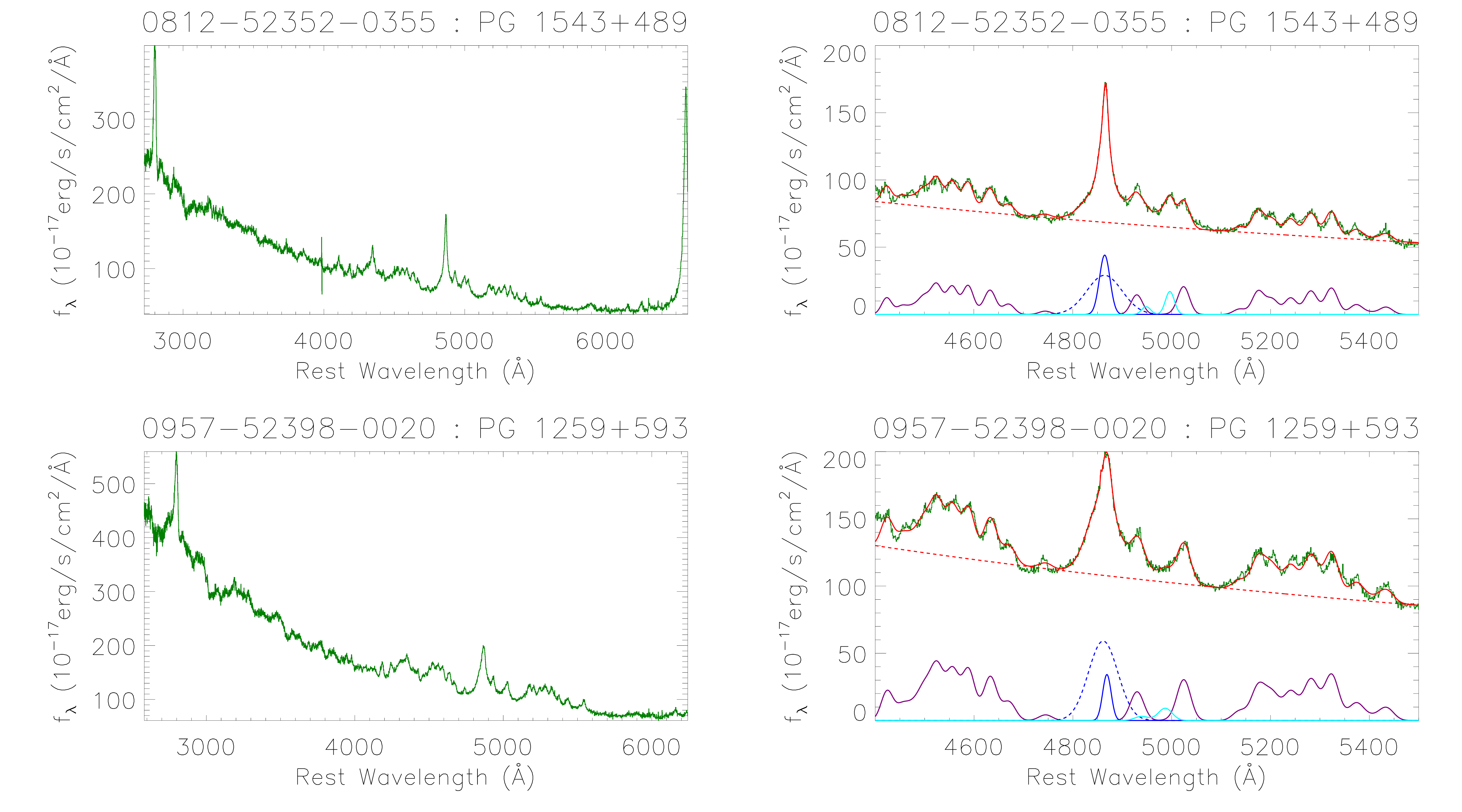}
\caption{Left panels show the SDSS spectra of PG 1543+489 and PG 1259+593. Right panels show the best fitting results (solid red 
lines) to the emission lines (in dark green) with rest wavelength from 4400\AA~ to 5600\AA. In each right panel, solid blue line 
shows the determined narrow H$\beta$, solid cyan line shows the determined [O~{\sc iii}]$\lambda4959,5007$\AA~ doublet, dashed red 
line shows the determined power law AGN continuum emission component underneath the emission lines, solid purple line shows the 
determined optical Fe~{\sc ii} emission features, dashed blue line shows the determined broad H$\beta$.}
\label{rep1}
\end{figure*}

	Before proceeding further, more discussions on the [N~{\sc ii}]$\lambda6548,6583$\AA~ and narrow H$\alpha$ are given as 
follows. Besides the model functions applied to the emission lines around H$\alpha$ as shown in Fig.~\ref{ha}, new model functions 
are applied without considerations of narrow [N~{\sc ii}] emissions, leading to the new determined $\chi^2/dof$ is to be 
$301.57/270\sim1.116$. Then, similar as what we have recently done in \citet{zm25}, the F-test technique can be applied to confirm 
that the confidence level is only around 0.87$\sigma$ to support the complicated model functions including not only broad components 
but also narrow component. In other words, through the SDSS provided spectrum, the narrow components are not preferred. But to show 
and discuss the not so reliable narrow [N~{\sc ii}] components could provide some clues to support the evolving NLRs scenario. 
Certainly, in the near future, spectrum with higher quality than the SDSS spectrum could provide more detailed information on the 
narrow [N~{\sc ii}]$\lambda6548,6583$\AA~ in SDSS J1251+0613.

\section{Necessary Discussions}

\subsection{comparing with previous results}
	
	In \citet{bg92}, there were three low redshift PG QSOs, PG 1259+593, PG 1543+489 and PG 2112+059, reported with equivalent 
width (EW) of [O~{\sc iii}] to be zero, indicating no apparent [O~{\sc iii}] emission lines. It is interesting to check whether such 
PG QSOs have similar emission line properties as SDSS J1251+0613.

	More interestingly, except PG 2112+059, high quality spectra can be collected from SDSS for the PG 1259+593 
(plate-mjd-fiberid=957-52398-20) and PG 1543+489 (plate-mjd-fiberid=812-52352-355) and shown in the left panels of Fig.~\ref{rep1}. 
Then, emission line fitting procedures are applied to describe the emission lines, especially the lines within the rest wavelength 
from 4400\AA~ to 5600\AA, including the H$\beta$, [O~{\sc iii}] and optical Fe~{\sc ii} emissions, as what we have done in SDSS 
J1251+0613. The best fitting results are shown in the right panels of Fig.~\ref{rep1}.

	Based on the best fitting results, at least two points can be confirmed. First, there are apparent narrow H$\beta$ emission 
lines (the component shown as solid blue line in each right panel of Fig.~\ref{rep1}) in PG 1259+593 and PG 1543+489. Second, there 
are weak but apparent [O~{\sc iii}]$\lambda5007$\AA~ emission lines (the component shown as solid cyan line in each right panel of 
Fig.~\ref{rep1}). Here, due to the determined second moment of the narrow H$\beta$ smaller than the second moment of 
[O~{\sc iii}]$\lambda5007$\AA, the determined narrow H$\beta$ is not considered as component from central BLRs but truly from NLRs. 
Moreover, after checking the database of SDSS quasars reported in \citet{sr11}, reliable line luminosities around $10^{42}$erg/s 
of narrow H$\beta$ have been reported in PG 1259+593 and PG 1543+489, to support the narrow H$\beta$. Therefore, probably due to 
high quality of SDSS spectrum leading to clearer determined emission components, apparent narrow emission lines of narrow H$\beta$ 
and weak but apparent [O~{\sc iii}]$\lambda4959,5007$ can be detected in both PG 1259+593 and PG 1543+489.

	Furthermore, after checking the spectroscopic results shown in Fig.~1 in \citet{bg92} for PG 1259+593 and 
PG 1543+489, we can find that the continuum emissions around 5100\AA~ are stronger in the SDSS spectra, indicating the 
PG 1259+593 and PG 1543+489 are brighter, when observed in SDSS. Therefore, the appearance of narrow [O~{\sc iii}] doublet in 
PG 1259+593 and PG 1543+489 in SDSS spectra is not due to weakened AGN emissions, but probably due to higher quality of SDSS spectra.

	Considering SDSS J1251+0613 having no apparent narrow Balmer emission lines nor narrow forbidden emission lines, the SDSS 
J1251+0613 is unique enough that the SDSS J1251+0613 is so-far the first broad line quasar without apparent narrow emission lines. 

\begin{figure}
\centering\includegraphics[width = 9cm,height=6cm]{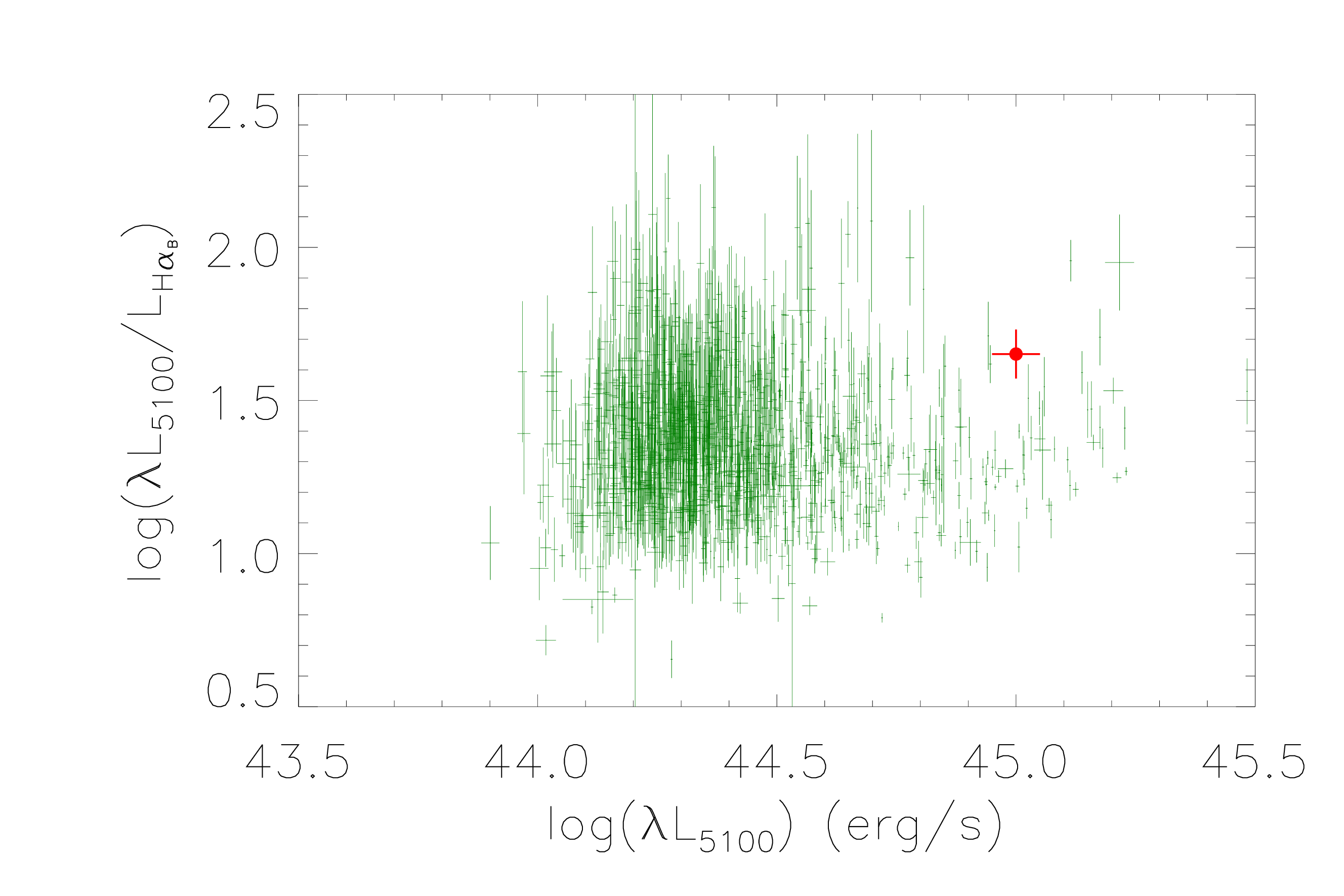}
\caption{On the dependence of $\log(\lambda L_{5100}/L_{H\alpha_B})$ on $\log(\lambda L_{5100})$ for the collected 1206 
normal SDSS quasars (dots plus error bars in dark green) and the SDSS J1251+0613 (solid circle plus error bars in red).}
\label{high}
\end{figure}

\begin{figure*}
\centering\includegraphics[width = 18cm,height=8cm]{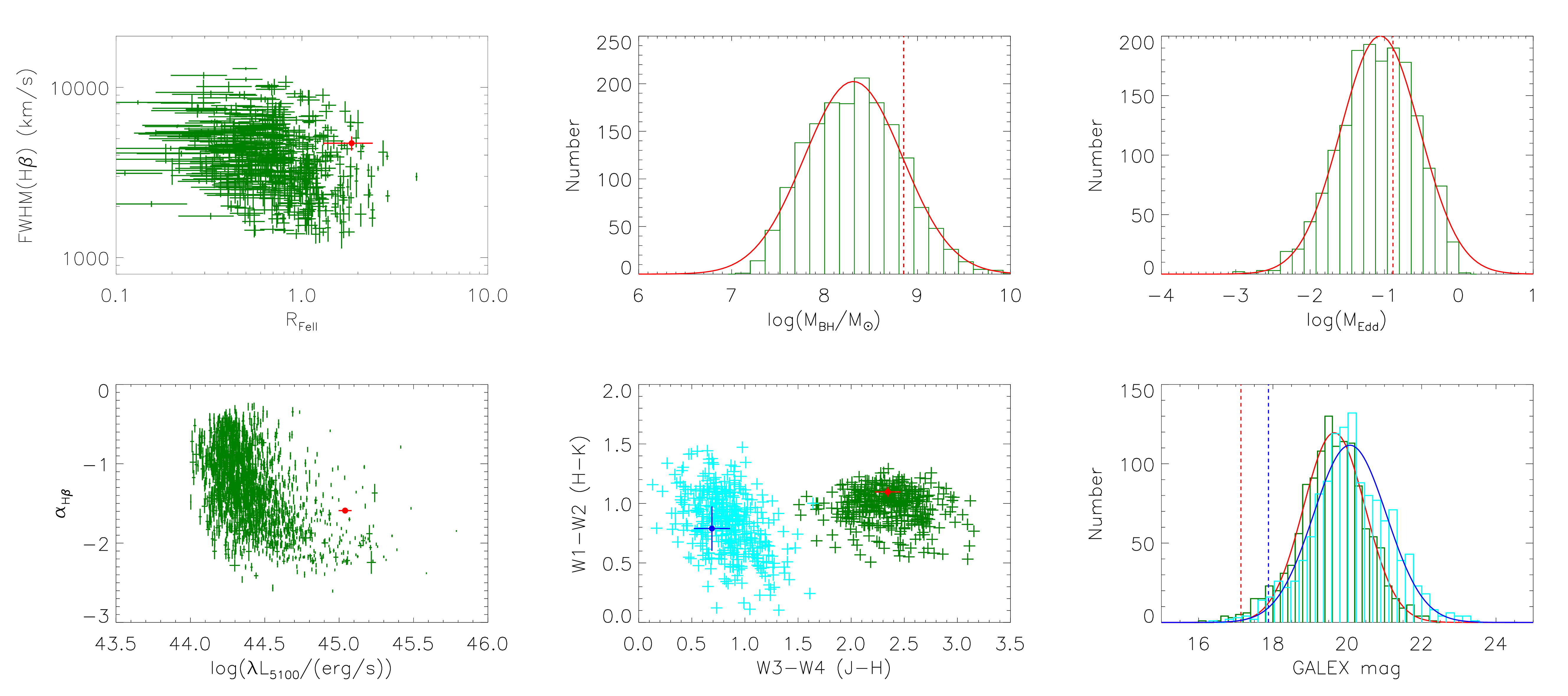}
\caption{Top left panel shows the dependence of FWHM(H$\beta$) on $R_{FeII}$ for the SDSS J1251+0613 (solid circle plus error bars 
in red) and the collected 576 SDSS normal quasars (smaller solid circles plus error bars in dark green). Top middle panel and top 
right panel show the BH mass distribution and Eddington ratio distributions of the collected 1642 SDSS normal quasars. In top 
middle panel and top right panel, solid line in red shows the Gaussian fitting results to the distribution, and vertical dashed 
red line marks the position for the SDSS J1251+0613. Bottom left panel shows the dependence of $\alpha_{H\beta}$ on continuum 
luminosity, with solid red circle for the SDSS J1251+0613. Bottom middle panel shows the IR color properties from the WISE (symbols 
in dark green) and from the 2MASS (symbols in cyan). In bottom middle panel, solid circle plus error bars in red and in blue show 
the WISE and 2MASS results for the SDSS J1251+0613. Bottom right panel shows the NUV (in dark green) and FUV (in cyan) band 
magnitude distributions, with solid line in red and in blue for the Gaussian fitting results to the distributions. In bottom right 
panel, vertical dashed line in red and in blue show the NUV and FUV magnitudes for the SDSS J1251+0613.}
\label{in}
\end{figure*}

\subsection{Other explanations for the lack of narrow emission lines?}

	There are no apparent narrow emission lines in the SDSS J1251+0613, probably due to strong obscurations on central NLRs. 
However, after considering the following point, effects of obscurations can be ruled out. Both the apparent blue continuum 
emissions and the apparent broad Balmer emission lines in SDSS J1251+0613 strongly indicate that the central regions are in the 
line-of-sight. Hence, once there were effects of obscurations on NLRs, the obscurations must have also effects on central BLRs 
in SDSS J1251+0613. However, through the reported line parameters in Table~1 for the broad H$\beta$ and broad H$\alpha$, the 
flux ratio is about $2.6_{-0.5}^{+0.6}$ of broad H$\alpha$ to broad H$\beta$, indicating there are no obscurations on central 
BLRs of SDSS J1251+0613. Therefore, effects of obscuration are not preferred to explain the lack of narrow emission lines in 
SDSS J1251+0613.

	There are no apparent narrow emission lines in the SDSS J1251+0613, probably due to central NLRs not covered by the SDSS 
fiber. For SDSS J1251+0613 at redshift 0.375, the SDSS fiber (3\arcsec in diameter) covered region has the radius about 24000pc, 
which is large enough to totally cover the NLRs of SDSS J1251+0613 with the continuum luminosity at 5100\AA~ about $10^{45}$erg/s 
leading to the expected NLRs size of about 8000pc to the central BH through the empirical relation \citep{liu13} between the 
NLRs size and the [O~{\sc iii}]$\lambda5007$\AA~ line luminosity. Here, due to no measurements of [O~{\sc iii}]$\lambda5007$\AA~ 
line luminosity in SDSS J1251+0613, the intrinsic correlation \citep{zh17, hb14, za03} has been applied between power-law AGN 
continuum luminosity and [O~{\sc iii}]$\lambda5007$\AA~ line luminosity. Therefore, the non-detected optical/NUV forbidden/permitted 
narrow emission lines in SDSS J1251+0613 are not due to restrictions of SDSS fiber sizes, unless the sizes in lateral dimension 
of NLRs were unexpectedly larger than the size in radial dimension of NLRs in SDSS J1251+0613.

	Moreover, as discussed in \citet{bl05}, very lower ionization parameter $U$ in NLRs could lead to very smaller EW of 
[O~{\sc iii}] emission line. There are two common ways leading to lower $U\propto\frac{1}{r_{NLRs}^2 n_e}$ in NLRs in the broad 
line quasar SDSS J1251+0613, very higher electron density $n_e$ in NLRs and/or very longer distance $R_{NLRs}$ from central BH. 
However, both higher $n_e$ and longer $R_{NLRs}$ could lead to stronger narrow H$\alpha$. Therefore, to explain the 
lack of narrow emission lines in SDSS J1251+0613, the dependence proposed in \citet{bl05} should be not preferred.

	Furthermore, no apparent narrow emission lines in the SDSS J1251+0613 could be probably explained after considering the 
SDSS spectrum at a very higher state could lead to non-detected narrow emission lines, due to very stronger continuum emissions 
and/or very strong broad emission lines overwhelming the narrow emission line features. However, based on the collected 1206 SDSS 
quasars with redshift between 0.35 and 0.4 (redshift 0.375 for SDSS J1251+0613) and with reliable continuum luminosity 
$\lambda L_{5100}$ at 5100\AA~ and reliable measurements of broad H$\alpha$ line luminosity $L_{H\alpha_B}$ in \citet{sr11}, 
Fig.~\ref{high} shows the dependence of luminosity ratio of $\lambda L_{5100}$ to $L_{H\alpha_B}$ on the continuum luminosity 
$\lambda L_{5100}$. Meanwhile, we also show the SDSS J1251+0613 in the figure. It is clear that the SDSS J1251+0613 has not very 
different properties from the normal quasars in the space of luminosity ratio of $\lambda L_{5100}$ to $L_{H\alpha_B}$ versus 
$\lambda L_{5100}$. In other words, the SDSS spectrum of SDSS J1251+0613 is not at a higher state with very stronger continuum 
emissions or very stronger broad line emissions. Therefore, the spectrum at a higher state cannot be applied to explain the 
disappearance of narrow emission lines in the SDSS J1251+0613.

\subsection{Comparing with normal broad line quasars}

	Based on the emission line properties of SDSS J1251+0613, the following parameters are determined and discussed, in order 
to check whether the SDSS J1251+0613 has unique properties different from normal broad line quasars.

	Due to the measured optical Fe~{\sc ii} emission features, the parameter $R_{FeII}\sim1.85\pm0.55$ can be estimated by the 
EW ratio of Fe~{\sc ii} features within rest wavelength from 4400\AA~ to 4700\AA~ to broad H$\beta$. Meanwhile, based on the 
determined profile of broad H$\beta$, the FWHM (full width at half maximum) of broad H$\beta$ is about 
FWHM(H$\beta$)$\sim$4800$\pm$532km/s. Then, based on the measured parameters at least 5 times larger than their corresponding 
uncertainties leading to the 576 SDSS quasars collected from \citet{sr11} with reliable $R_{Feii}$ and reliable FWHM of broad 
H$\beta$ and with redshift between 0.35 and 0.4 (redshift 0.375 for SDSS J1251+0613), top left panel of Fig.~\ref{in} shows the 
dependence of FWHM(H$\beta$) on $R_{FeII}$ for the SDSS J1251+0613 and the collected 576 normal SDSS quasars. There are 
no strong clues to support that SDSS J1251+0613 behaves very differently from the normal quasars in the space of 
FWHM(H$\beta$) versus $R_{FeII}$.

	Due to the measured broad Balmer emission lines and the continuum luminosity at 5100\AA, through the virialization 
assumption accepted to the broad line clouds in BLRs \citep{pf04, gh05, vp06, sr11}, the virial BH mass $M_{BH}$ and the 
corresponding Eddington ratio $M_{Edd}$ of SDSS J1251+0613 can be estimated as
\begin{equation}
\begin{split}
	\log(\frac{M_{BH}}{\rm M_\odot})&= 0.91+0.5\log(\frac{\lambda L_{5100}}
	{\rm 10^{44}erg/s}) + 2\log(\frac{FWHM(H\beta)}{\rm km/s})\\
	&\sim~8.85 \\
	M_{Edd}&= \frac{L_{bol}}{1.38\times10^{38}M_{BH}/{\rm M_\odot}}=
	\frac{9.26\lambda L_{5100}}{1.38\times10^{38}M_{BH}/{\rm M_\odot}}\\
	&\sim~0.13
\end{split}
\end{equation}.
Here, in order to compare the BH mass calculated by the same equation and then the corresponding Eddington ratio between the SDSS 
J1251+0613 and the normal SDSS quasars, the equation (2) and (5) and the bolometric correction factor 9.26 applied in \citet{sr11} 
have been accepted. Then, based on measured parameters at least 5 times larger than their corresponding uncertainties leading to 
the 1642 SDSS quasars collected from \citet{sr11} with reliable $M_{BH}$ and reliable bolometric luminosity and with redshift 
between 0.35 and 0.4, top middle and top right panels of Fig.~\ref{in} show the distributions of $M_{BH}$ and $M_{Edd}$ for the 
normal SDSS quasars. There are no strong clues to support SDSS J1251+0613 having very different BH mass or Eddington ratio from 
those of the normal quasars.

	Furthermore, based on the measured power law continuum emissions in the SDSS J1251+0613, the determined slope of the power 
law component is about $\alpha_{H\beta}\sim$-1.59$\pm$0.03, and the determined continuum luminosity at 5100\AA~ is about $10^{45}$erg/s. 
Then, based on the measured parameters at least 5 times larger than their corresponding uncertainties leading to the 1510 SDSS 
quasars collected from \citet{sr11} with reliable $\alpha_{H\beta}$ and reliable continuum luminosity at 5100\AA~ ($\lambda L_{5100}$) 
and with redshift between 0.35 and 0.4, bottom left panel of Fig.~\ref{in} shows the dependence of $\alpha_{H\beta}$ on continuum 
luminosity for the 1510 normal SDSS quasars and the SDSS J1251+0613. It looks like that the SDSS J1251+0613 has a bit 
larger $\alpha_{H\beta}$ for given $\lambda L_{5100}$, however, after checking the SDSS quasars near to the SDSS J1251+0613 in the 
space of $\alpha_{H\beta}$ versus $\lambda L_{5100}$, there are apparent narrow emission lines in the SDSS quasars. Therefore, there 
should not probably strong clues to support SDSS J1251+0613 being very different from the normal quasars in the space of 
$\alpha_{H\beta}$ versus $\lambda L_{5100}$.

	Moreover, after collected magnitudes from the Wide-field Infrared Survey Explorer (WISE) and the Two Micron All-Sky Survey 
(2MASS) for the SDSS J1251+0613 and for the 376 normal SDSS quasars collected from \citet{sr11} with redshift between 0.35 and 0.4, 
bottom middle panel shows the Infrared color properties. Meanwhile, after collected magnitudes from the Galaxy Evolution Explorer 
(GALEX) for the SDSS J1251+0613 and for the 1283 normal SDSS quasars collected from \citet{sr11} with redshift between 0.35 and 0.4, 
bottom right panel shows the distributions of near-UV (NUV) and far-UV (FUV) band magnitudes. It is clear that there are no strong 
clues to support SDSS J1251+0613 being very different from the normal quasars through both IR and UV band properties.

	Therefore, at the current stage, based on the physical properties of central regions, there are no clues to support SDSS 
J1251+0613 being different from the normal SDSS quasars. 

\begin{figure}
\centering\includegraphics[width = 8cm,height=5cm]{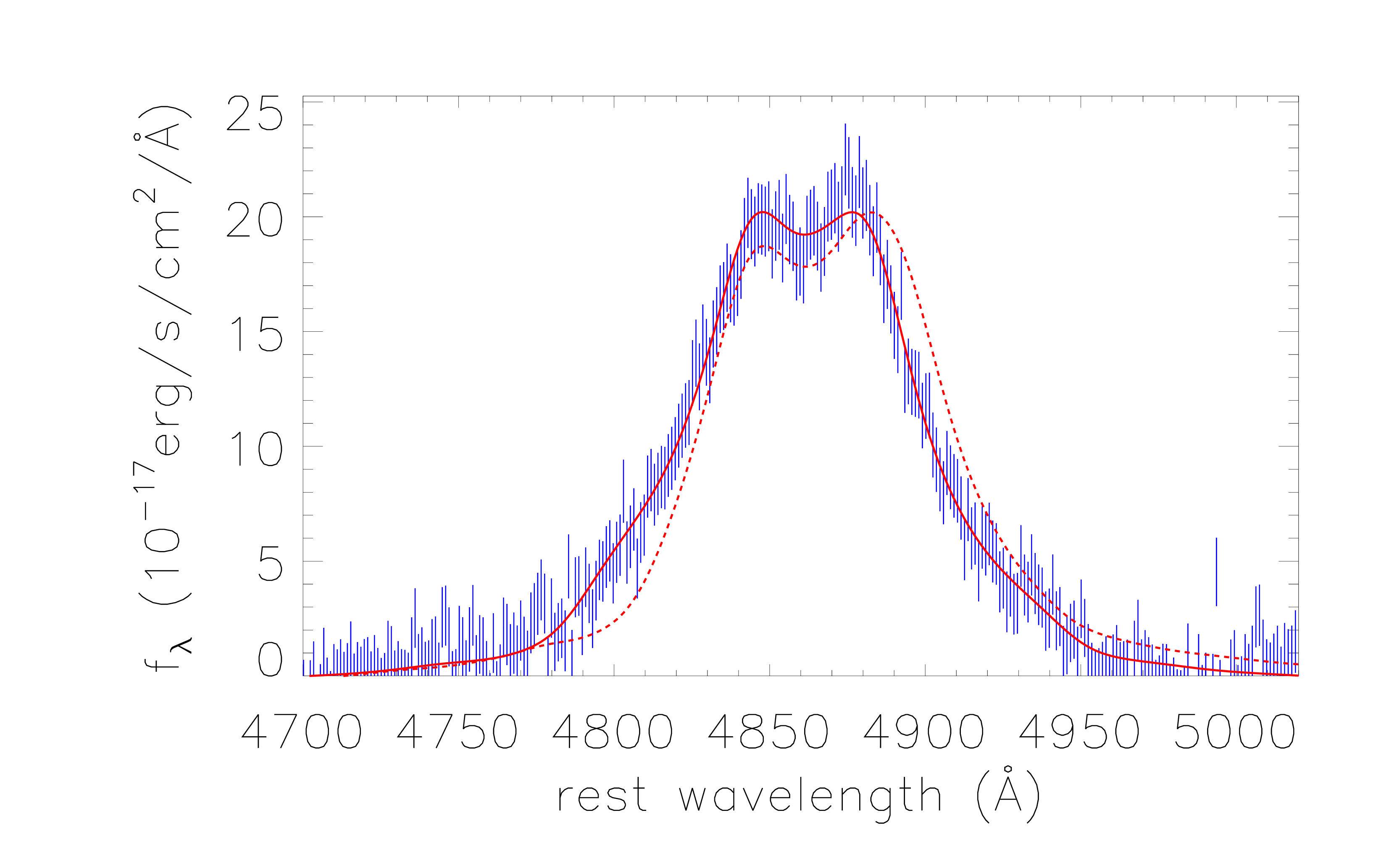}
\caption{The best descriptions (solid red line) to the asymmetric broad H$\beta$ by the elliptical accretion disk model. Dots plus 
error bars show the broad H$\beta$ after removing the optical Fe~{\sc ii} emission features and the AGN continuum emissions. Dashed 
red line shows the expected line profile of broad H$\beta$ around 2030, under the assumption of disk-like BLRs lying into central 
accretion disk in SDSS J1251+0613.}
\label{dbp}
\end{figure}

\subsection{Disk origin for the broad Balmer emission lines?}

	In Fig.~\ref{hb} and Fig.~\ref{ha}, asymmetric broad Balmer emission lines can be found. We have accepted the asymmetric 
broad Balmer emission lines as clues to support radial outflows in BLRs in SDSS J1251+0613, to support our hypothesis of  
evolving NLRs model. Here, besides the radial outflows, accretion disk origin (disk-like BLRs lying into central accretion disk) 
can be also applied to describe the asymmetric broad Balmer emission lines, such as the discussed results in \citet{el95, ss03, 
st03, ss17, ms24, wg24}, to provide further clues to distinguish the outflowing model and the accretion disk origin for the 
asymmetric Balmer emission lines in the near future.

	In the subsection, we employ the elliptical accretion disk model proposed in \citet{el95}, which has also been 
applied in our recent work in \citet{zh24c, zh24d, zh25a}. There are not detailed discussions on the accretion disk model any more, but 
simple information are given on the following seven model parameters, the inner radius $r_0$ (in units of $R_{G}$ with $R_G$ as the 
Schwarzschild radius) and the outer radius $r_1$ (in units of $R_{G}$) of the emission region, the inclination angle $i$ of the 
emission region, the line emissivity power-law index $f_r\propto r^{-q}$, the local turbulent broadening parameter $\sigma_L$ (in 
units of km/s), the eccentricity $e$ of the emission region, and the orientation angle $\phi_0$. Then, through the Levenberg-Marquardt 
least-squares minimization technique (the known MPFIT package) \citep{mc09}, the best fitting results to the broad H$\beta$ (after 
removing the optical Fe~{\sc ii}  emission features and the AGN continuum emissions) can be determined and shown in Fig.~\ref{dbp} 
with $\chi^2/dof\sim0.65$. The determined model parameters are listed in Table~1. Similar results can be determined to the asymmetric 
broad H$\alpha$, here we do not show the corresponding results related to broad H$\alpha$ any more.

	Based on the determined model parameters, the probable precession period \citep{st03} related to the inner emission regions 
of the assumed disk-like BLRs in SDSS J1251+0613 is about 
\begin{equation}
\begin{split}
	T_{pre}&\sim10.4\frac{M_{BH}}{10^6{\rm M_\odot}}\frac{1+e}{(1-e)^{1.5}}(\frac{r_{0}}{10^3{\rm R_G}})^{2.5} {\rm yrs} \\
	&\sim10.4\times~708\times~\frac{1.87}{0.047}\times~(0.096)^{2.5}\sim840{\rm yrs}
\end{split}
\end{equation}. 
However, considering the eccentricity being 0.8, a very small difference in $\phi_0$ can lead to apparent variability of line profile. 
As the shown dashed red line in Fig.~\ref{dbp} with time duration about 23years from MJD=54270, about 10\degr difference in $\phi_0$ 
related to inner emission regions can lead to different line profile from that observed at MJD=54270. In other words, if new 
spectroscopic results can be obtained around 2030, expected different line profile from that observed at MJD=54270 should indicate 
accretion disk origin rather than the radial outflows will be preferred in SDSS J1251+0613. At the current stage, we cannot give a 
strong conclusion on the origin of the asymmetric features in broad Balmer lines.

\section{Conclusions}

	In this work, we provide the first clear report on non-detected forbidden/permitted narrow emission lines in 
the blue quasar SDSS J1251+0613, providing potential evidence on materials in AGN NLRs coming from central high-density material 
clouds in normal BLRs through central high velocity radial flows. Broad absorption lines can be used as good indicators of central 
high velocity radial flows. However, in SDSS J1251+0613, we have checked properties of the broad low-ionization emission lines of 
Mg~{\sc ii} and Balmer lines, and found no apparent broad absorption features. Probable high-ionization emission lines especially 
in the UV band, such as C~{\sc iv} line, should provide further clues on high velocity radial flows from central regions in SDSS 
J1251+0613. At current stage, there are no strong clues to support that the SDSS J1251+0613 have unique properties very different 
from the normal SDSS low redshift quasars, and no strong clues to confirm the asymmetric broad Balmer emission lines related to 
radial outflows rather than coming from disk-like BLRs lying into central accretion disk. To detect and report a sample of such blue 
quasars (BLAGNs) without narrow emission lines could provide further clues to support the proposed evolving AGN NLRs in the near future.

\begin{acknowledgements}
Zhang gratefully acknowledge the anonymous referee for giving us constructive comments and suggestions to greatly 
improve the paper. Zhang gratefully thanks the kind financial support from GuangXi University and the kind grant support from 
NSFC-12173020 and NSFC-12373014, and the support from the Guangxi Talent Programme (Highland of Innovation Talents). This 
manuscript has made use of the data from the SDSS projects. The SDSS-III web site is http://www.sdss3.org/. SDSS-III is managed 
by the Astrophysical Research Consortium for the Participating Institutions of the SDSS-III Collaboration.
\end{acknowledgements}


\begin{thebibliography}{   }
\bibitem[\protect\citeauthoryear{Alam et al.}{2015}]{as15} 
Alam, S.; Albareti, F. D.; Allende Prieto, C.; et al., 2015, ApJS, 219, 12
\bibitem[\protect\citeauthoryear{Antonucci}{1993}]{ar93} 
Antonucci, R., 1993, ARA\&A, 31, 473
\bibitem[\protect\citeauthoryear{Baskin \& Laor}{2005}]{bl05} 
Baskin, A.; Laor, A., 2005, MNRAS, 358, 1043
\bibitem[\protect\citeauthoryear{Bentz et al.}{2009}]{bw09} 
Bentz, M. C.; Walsh, J. L.; Barth, A. J.; et al., 2009, ApJ, 705, 199
\bibitem[\protect\citeauthoryear{Bentz et al.}{2013}]{ben13} 
Bentz, M. C.; Denney, K. D.; Grier, C. J.; et al., 2013, ApJ, 779, 109
\bibitem[\protect\citeauthoryear{Boroson \& Green}{1992}]{bg92} 
Boroson, T. A.; Green, R. F., 1992, ApJS, 80, 109
\bibitem[\protect\citeauthoryear{Brotherton et al.}{2001}]{bt01} 
Brotherton, M. S.; Tran, H. D.; Becker, R. H.; Gregg, M. D.; Laurent-Muehleisen, S. A.; White, R. L., 2001, ApJ, 546, 775
\bibitem[\protect\citeauthoryear{Cao}{2010}]{ca10} 
Cao, X. W., 2010, ApJ, 724, 855
\bibitem[\protect\citeauthoryear{Costa-Souza et al.}{2024}]{cr24} 
Costa-Souza, J. H.; Riffel, R. A.; Dors, O. L.; Riffel, R.; da Rocha-Poppe, P. C, 2024, MNRAS, 527, 9192
\bibitem[\protect\citeauthoryear{Dempsey \& Zakamska}{2018}]{dz18} 
Dempsey, R.; Zakamska, N. L., 2018, MNRAS, 477, 4615
\bibitem[\protect\citeauthoryear{Du et al.}{2014}]{dw14} 
Du, P.; Wang, J. M.; Hu, C.; Valls-Gabaud, D.; Baldwin, J. A.; Ge, J.; Xue, S., 2014, MNRAS, 438, 2828
\bibitem[\protect\citeauthoryear{Elitzur \& Ho}{2009}]{eh09} 
Elitzur, M.; Ho, L. C., 2009, ApJ, 701, L91
\bibitem[\protect\citeauthoryear{Elitzur \& Netzer}{2016}]{en16} 
Elitzur, M.; Netzer, H., 2016, MNRAS, 459, 585
\bibitem[\protect\citeauthoryear{Eracleous et al.}{1995}]{el95}
Eracleous, M.; Livio, M.; Halpern, J. P.; Storchi-Bergmann, T., 1995, ApJ, 438, 610
\bibitem[\protect\citeauthoryear{Fischer et al.}{2013}]{fc13} 
Fischer, T. C.; Crenshaw, D. M.; Kraemer, S. B.; Schmitt, H. R., 2013, ApJS, 209, 1
\bibitem[\protect\citeauthoryear{Fischer et al.}{2018}]{fk18} 
Fischer, T. C.; Kraemer, S. B.; Schmitt, H. R.; et al., 2018, ApJ, 856, 102
\bibitem[\protect\citeauthoryear{Greene \& Ho}{2005}]{gh05} 
Greene, J. E.; Ho, L. C., 2005, ApJ, 630, 122
\bibitem[\protect\citeauthoryear{Hawkins}{2004}]{hm04} 
Hawkins, M. R. S., 2004, A\&A, 424, 519
\bibitem[\protect\citeauthoryear{Heckman \& Best}{2004}]{hb14} 
Heckman, T. M.; Best, P. N., 2014, ARA\&A, 52, 589
\bibitem[\protect\citeauthoryear{Ho et al.}{2012}]{hk12} 
Ho, L. C.; Kim, M.; Terashima, Y., 2012, ApJL, 759, L16
\bibitem[\protect\citeauthoryear{Holden \& Tadhunter}{2025}]{ht25} 
Holden, L. R.; Tadhunter, C. N., 2025, MNRAS, 536, 1857
\bibitem[\protect\citeauthoryear{Ichikawa et al.}{2015}]{ip15} 
Ichikawa, K.; Packham, C.; Almeida, C. R.; et al., 2015, ApJ, 803, 57
\bibitem[\protect\citeauthoryear{Kaspi et al.}{2000}]{ks00} 
Kaspi, S.; Smith, P. S.; Netzer, H.; Maoz, D.; Jannuzi, B. T.; Giveon, U., 2000, ApJ, 533, 631
\bibitem[\protect\citeauthoryear{Kovacevic}{2010}]{kp10} 
Kovacevic, J.; Popovic, L. C.; Dimitrijevic, M. S., 2010, ApJS, 189, 15
\bibitem[\protect\citeauthoryear{Kuzmicz et al.}{2021}]{ks21} 
Kuzmicz, A.; Sethi, S.; Jamrozy, M., 2021, ApJ, 922, 52
\bibitem[\protect\citeauthoryear{Liu et al.}{2013}]{liu13} 
Liu, G.; Zakamska, N. L.; Greene, J. E.; Nesvadba, N. P. H.; Liu, X., 2013, MNRAS, 430, 2327,
\bibitem[\protect\citeauthoryear{Liu et al.}{2019}]{ll19} 
Liu, H.; Liu, W.; Dong, X.; Zhou, H.; Wang, T.; Lu, H.; Yuan, W., 2019, ApJS, 243, 21
\bibitem[\protect\citeauthoryear{Markwardt}{2009}]{mc09} 
Markwardt, C. B., 2009, ASPC, 411, 251
\bibitem[\protect\citeauthoryear{Marsango et al.}{2024}]{ms24}
Marsango, D.; Schimoia, J. S.; Rembold, S. B.; Storchi-Bergmann, T.; Nemmen, R. S.; Couto, G. S.; Souza-Oliveira, G.; 
Costa-Souza, J., 2024, MNRAS, 529, 3089
\bibitem[\protect\citeauthoryear{Molina et al.}{2022}]{mh22} 
Molina, J.; Ho, L. C.; Wang, R.; et al., 2022, ApJ, 935, 72
\bibitem[\protect\citeauthoryear{Netzer}{2015}]{nh15} 
Netzer, H., 2015, ARA\&A, 53, 365
\bibitem[\protect\citeauthoryear{Oh et al.}{2015}]{oy15} 
Oh, K.; Yi, S. K.; Schawinski, K.; Koss, M.; Trakhtenbrot, B.; Soto, K., 2015, ApJS, 219, 1
\bibitem[\protect\citeauthoryear{Park et al.}{2012}]{pw12} 
Park, D.; Woo, J. H.; Treu, T.; et al., 2012, ApJ, 747, 30
\bibitem[\protect\citeauthoryear{Peterson et al.}{2004}]{pf04} 
Peterson, B. M.; Ferrarese, L.; Gilbert, K. M.; et al., 2004, ApJ, 613, 682
\bibitem[\protect\citeauthoryear{Pol \& Wadadekar}{2017}]{pw17} 
Pol, N.; Wadadekar, Y., 2017, MNRAS, 465, 95 
\bibitem[\protect\citeauthoryear{Pons \& Watson}{2016}]{pw16} 
Pons, E.; Watson M. G., 2016, A\&A, 594, A72
\bibitem[\protect\citeauthoryear{Schimoia et al.}{2017}]{ss17} 
Schimoia, J. S. ; Storchi-Bergmann, T.; Winge, C.; Nemmen, R. S.; Eracleous, M., 2017, MNRAS, 472, 2170
\bibitem[\protect\citeauthoryear{Selsing et al.}{2016}]{sf16} 
Selsing, J.; Fynbo, J. P. U.; Christensen, L.; Krogager, J. K., 2016, A\&A, 585, 87
\bibitem[\protect\citeauthoryear{Shen et al.}{2011}]{sr11} 
Shen, Y.; Richards, G. T.; Strauss, M. A.; et al., 2011, ApJS, 194, 45
\bibitem[\protect\citeauthoryear{Shull et al.}{2012}]{ss12} 
Shull, J. M.; Stevans, M.; Danforth, C. W.. 2012, ApJ, 752, 162
\bibitem[\protect\citeauthoryear{Sigut \& Pradhan}{2003}]{sp03} 
Sigut, T. A. A.; Pradhan, A. K., 2003, ApJS, 145, 15
\bibitem[\protect\citeauthoryear{Smee et al.}{2013}]{ss13} 
Smee, S. A.; Gunn, J. E.; Uomoto, A.; et al., 2013, AJ, 146, 32
\bibitem[\protect\citeauthoryear{Speranza et al.}{2024}]{sr24} 
Speranza, G.; Ramos Almeida, C.; Acosta-Pulido, J. A.; et al., 2024, A\&A, 681, 63
\bibitem[\protect\citeauthoryear{Strateva et al.}{2003}]{ss03}
Strateva, I. V., Strauss, M. A., Hao, L., et al. 2003, AJ, 126, 1720
\bibitem[\protect\citeauthoryear{Storchi-Bergmann et al.}{2003}]{st03}
Storchi-Bergmann, T.; Nemmen da Silva, R., Eracleous, M., et al., 2003, ApJ, 598, 956
\bibitem[\protect\citeauthoryear{Sulentic et al.}{2000}]{sm00} 
Sulentic, J. W.; Marziani, P.; Dultzin-Hacyan, D., 2000, ARA\&A, 38, 521
\bibitem[\protect\citeauthoryear{Tran}{2001}]{tr01} 
Tran, H. D., 2001, ApJ, 554, L19
\bibitem[\protect\citeauthoryear{Tran}{2003}]{tr03} 
Tran, H. D., 2003, ApJ, 583, 632
\bibitem[\protect\citeauthoryear{Trindade Falcao et al.}{2021}]{tk21} 
Trindade Falcao, A.; Kraemer, S. B.; Fischer, T. C.; et a;., 2021, MNRAS, 500, 1491
\bibitem[\protect\citeauthoryear{Trefoloni et al.}{2024}]{tl24} 
Trefoloni, B.; Lusso, E.; Nardini, E.; et al., 2024, A\&A, 689, 109
\bibitem[\protect\citeauthoryear{Vanden Berk et al.}{2001}]{vr01} 
Vanden Berk, D. E.; Richards, G. T.; Bauer, A.; et al., 2001, AJ, 122, 549
\bibitem[\protect\citeauthoryear{Vestergaard \& Peterson}{2006}]{vp06}
Vestergaard, M., Peterson, B. M. 2006, ApJ, 641, 689
\bibitem[\protect\citeauthoryear{Ward et al.}{2024}]{wg24} 
Ward, C.; Gezari, S.; Nugent, P.; 2024, ApJ, 961, 172
\bibitem[\protect\citeauthoryear{Woo et al.}{2015}]{wy15} 
Woo, J. H.; Yoon, Y.; Park, S.; Park, D.; Kim, S. C., 2015, ApJ, 801, 38
\bibitem[\protect\citeauthoryear{Zakamska}{2003}]{za03} 
Zakamska, N. L.; Strauss, M. A.; Krolik, J. H.; et al., 2003, AJ, 126, 2125
\bibitem[\protect\citeauthoryear{Zhang}{2014}]{zh14} 
Zhang, X. G., 2014, MNRAS, 438, 557
\bibitem[\protect\citeauthoryear{Zhang \& Feng}{2017}]{zh17} 
Zhang, X. G.; Feng, L. L., 2017, MNRAS, 468, 620
\bibitem[\protect\citeauthoryear{Zhang}{2021}]{zh21} 
Zhang, X. G., 2021, ApJ, 909, 16
\bibitem[\protect\citeauthoryear{Zhang}{2022a}]{zh22} 
Zhang, X. G., 2022a, ApJS, 260, 31
\bibitem[\protect\citeauthoryear{Zhang \& Zhao}{2022b}]{zh22b} 
Zhang, X. G.; Zhao S., 2022b, ApJ, 937, 105
\bibitem[\protect\citeauthoryear{Zhang}{2023}]{zh23} 
Zhang, X. G., 2023, ApJS, 267, 36
\bibitem[\protect\citeauthoryear{Zhang}{2024a}]{zh24} 
Zhang, X. G., 2024a, ApJ, 960, 108
\bibitem[\protect\citeauthoryear{Zhang}{2024b}]{zh24b} 
Zhang, X. G., 2024b, MNRAS, 530, 4346
\bibitem[\protect\citeauthoryear{Zhang}{2024c}]{zh24c} 
Zhang, X. G., 2024c, MNRAS Letters, 529, 169, arXiv:2402.04861
\bibitem[\protect\citeauthoryear{Zhang}{2024d}]{zh24d} 
Zhang, X. G., 2024d, MNRAS Letters, 529, 41, arXiv:2401.17683
\bibitem[\protect\citeauthoryear{Zhang}{2025a}]{zh25a} 
Zhang, X. G., 2025, A\&A Letters, 696, 22, arXiv:2504.05531
\bibitem[\protect\citeauthoryear{Zheng et al.}{2025}]{zm25} 
Zheng, Q.; Ma, Y. S.; Zhang, X. G.; Yuan, Q. R., Bian W. H., 2025, ApJS acceptd, arXiv:2502.15299
\bibitem[\protect\citeauthoryear{Zheng et al.}{1990}]{zs90} 
Zheng, W.; Binette, L.; Sulentic, J. W., 1990, ApJ, 365, 115
\bibitem[\protect\citeauthoryear{Zheng et al.}{1997}]{zk97}  
Zheng, W.; Kriss, G. A.; Telfer, R. C.; Grimes, J. P.; Davidsen, A. F., 1997, ApJ, 339, 742
\end{thebibliography}
\end{document}